\documentclass[%
 amsmath,amssymb,
 aps,
groupedaddress,
superscriptaddress,
]{revtex4-2}

\usepackage{graphicx}
\usepackage{mathrsfs}
\usepackage{amsmath}
\usepackage{placeins}
\usepackage[export]{adjustbox}
\usepackage[abs]{overpic}
\usepackage{subfig}
\usepackage[%
 colorlinks=true,
 pdfborder={0 0 0},
 linkcolor=blue,
 citecolor=blue,
 urlcolor=blue
]{hyperref}
\usepackage{placeins}
\usepackage{tikz}
\usepackage{dcolumn}
\usepackage{bm}

\begin{document}


\title[]{Is vortex stretching the main cause of the turbulent energy cascade?}
	\author{Maurizio Carbone}
	\affiliation{Dipartimento di Ingegneria Meccanica e Aerospaziale, Politecnico di Torino, Corso Duca degli Abruzzi 24, 10129 Torino, Italy}
	\affiliation{Department of Civil and Environmental Engineering, Duke University, Durham, North Carolina 27708, USA}
	\author{Andrew D. Bragg}\email{andrew.bragg@duke.edu}
	\affiliation{Department of Civil and Environmental Engineering, Duke University, Durham, North Carolina 27708, USA}

\date{\today}

\begin{abstract}
In three dimensional turbulence there is on average a cascade of kinetic energy from the largest to the smallest scales of the flow. While the dominant idea is that the cascade occurs through the physical process of vortex stretching, evidence for this is debated. In the framework of the Karman-Howarth equation for the two point turbulent kinetic energy, we derive a new result for the average flux of kinetic energy between two points in the flow that reveals the role of vortex stretching. However, the result shows that vortex stretching is in fact not the main contributor to the average energy cascade; the main contributor is the self-amplification of the strain-rate field. We emphasize the need to correctly distinguish and not conflate the roles of vortex stretching and strain-self amplification in order to correctly understand the physics of the cascade, and also resolve a paradox regarding the differing role of vortex stretching on the mechanisms of the energy cascade and energy dissipation rate. Direct numerical simulations are used to confirm the results, as well as provide further results and insights on vortex stretching and strain-self amplification at different scales in the flow. Interestingly, the results imply that while vortex stretching plays a sub-leading role in the average cascade, it may play a leading order role during large fluctuations of the energy cascade about its average behavior.
\end{abstract}

\maketitle

\section{Introduction}

Turbulence in a classical fluid, governed by the incompressible Navier-Stokes equation (NSE), is a paradigmatic example of a high-dimensional system that exists in a state far from thermodynamic equilibrium. In three dimensions (3D), it exhibits on average a cascade of energy from the largest scales of the system, where the energy is injected, to the smallest scales, where it is dissipated by the action of viscous stresses \cite{falkovich09}. While the cause of this cascade ultimately arises from inertial forces in the fluid, a detailed understanding of the mechanism(s) driving this cascade remains elusive, and continues to be an important area of investigation \cite{ballouz18}. \footnote[1]{
Strictly speaking, the flux of kinetic energy through the turbulent flow scales can only be strictly considered as a cascade if the flux is constant and if the mechanism transferring the energy from one scale to another is sufficiently local. Following standard (though imprecise) terminology, we shall refer to the nonlinear flux of energy across scales as a cascade, irrespective of whether the flux is really constant and local.}


Richardson proposed \cite{richardson} that the cascade occurs through a hierarchical process of instabilities whereby eddies break down and pass their energy to smaller eddies. However, there is no clear connection between this mechanism and the underlying NSE. An alternative idea, that has become the dominant paradigm, is that the stretching of vorticity drives the energy cascade \cite{taylor32,taylor38,tennekes,davidson,doan18}. Since the process of vortex stretching is a feature of the underlying dynamical equations \cite{pope}, this is an appealing candidate for the energy cascade mechanism. However, theoretical demonstrations of the direct link between vortex stretching and the energy cascade are limited. Perhaps the main result on this was derived in \cite{borue98}, where a closure model was used to obtain a result relating the energy cascade and vortex stretching. Another result is that derived in \cite{davidson} for the limit of the small turbulent scales that appears to reveal a connection between vortex stretching and the skewness of the longitudinal fluid velocity increments (which is related to the cascade of energy through the turbulent flow scales). Numerical studies have reported evidence that appears consistent with the idea that vortex stretching drives the energy cascade \cite{davidson08,doan18}. However, it is possible that these numerical results only reflect correlations between the quantities, not causal connections, and/or that vortex stretching is part, but not the sole mechanism. A clear demonstration of any such causal connection requires theoretical insight. Moreover, theoretical problems with the vortex stretching mechanism have also been discussed in the literature. For example, in \cite{tsinober} and \cite{sagaut18} it is argued that vortex stretching hinders the fluid kinetic energy dissipation, and that this implies that vortex stretching hinders the energy cascade, since dissipation is supposed to be the end result of the cascade. 

In this paper, we seek to resolve these issues by means of theoretical and numerical analysis, and careful argumentation. The outline of the paper is as follows: In \S\ref{PMEC}, using a Karman-Howarth type equation for the two point turbulent kinetic energy, we derive a new result for the flux of kinetic energy between two points in the flow that reveals the role of vortex stretching, as well as that of the self-amplification of the strain-rate field. We then provide a thorough and careful discussion of the correct physical interpretation of the result and the roles of vortex stretching and self-amplification of the strain-rate field in the energy cascade. In \S\ref{EDEC}, we then discuss the connection between the energy dissipation and energy cascade in turbulence, and provide an argument to show that the claim in \cite{tsinober} and \cite{sagaut18} that vortex stretching hinders the energy cascade is incorrect. In \S\ref{NRD} we present results from Direct Numerical Simulations that confirm the analytical results from \S\ref{PMEC}, and also provide insight into the roles of vortex stretching and strain self-amplification at different scales in the flow and during fluctuations of the energy cascade about its average behavior.

\section{Physical mechanisms of the energy cascade}\label{PMEC}

A traditional way to analyze the multiscale properties of turbulence is through the velocity increments $\Delta\bm{u}(\bm{x},\bm{r},t)\equiv \bm{u}(\bm{x}+\bm{r}/2,t)-\bm{u}(\bm{x}-\bm{r}/2,t)$, where $\bm{u}$ is the fluid velocity, $\bm{x}$ is a point in the flow, and $\bm{r}$ is the vector separating two points in the flow \cite{kolmogorov41a,pope,frisch}. For statistically homogeneous turbulence, the equation for $\mathcal{K}(\bm{r},t)\equiv\langle \|\Delta\bm{u}(\bm{r},t)\|^2\rangle/2$, the ensemble averaged turbulent kinetic energy (per unit mass) at scale $r\equiv\|\bm{r}\|$, is 
\begin{eqnarray}
\partial_t\mathcal{K}=-\bm{\partial_r\cdot}\bm{T} +2\nu\bm{\partial_r}^2\mathcal{K}-2\langle\epsilon\rangle+W,\label{dtK}
\end{eqnarray}
which is essentially the Karman-Howarth equation \cite{karman38,hill01}. In this equation, $\bm{\partial_r\cdot T}\equiv(1/2)\bm{\partial_r\cdot}\langle\|\Delta\bm{u}\|^2\Delta\bm{u}\rangle$ is the the nonlinear energy flux, $\nu$ is the fluid kinematic viscosity, $\langle\epsilon\rangle$ is the average kinetic energy dissipation rate, and $W$ represents energy injection into the flow. For statistically stationary 3D turbulence, if $W$ only acts at the large scales $L$, then in the inertial range ${\eta\ll r\ll L}$ (where $\eta$ is the Kolmogorov length scale \cite{pope}), $\bm{\partial_r\cdot}\bm{T}=-2\langle\epsilon\rangle$, corresponding to a constant downscale cascade of energy. 

\subsection{Expression for the energy cascade} 
While it is usually thought that the energy cascade described by $\bm{\partial_r\cdot T}$ is driven by vortex stretching, there is no explicit or obvious mathematical connection between the two. However, we now derive a result that shows how $\bm{\partial_r\cdot T}$ is related to the stretching of vorticity at scale $r$, which may be considered using the velocity gradient filtered at scale $r$. Our result is for isotropic turbulence, although it may apply approximately to more general flows under Kolmogorov's hypothesis of small-scale isotropy of the turbulence \cite{pope}.

We first introduce $\bm{u}=\widetilde{\bm{u}}+\bm{u}'$, where $\widetilde{\bm{u}}(\bm{x},t)\equiv\int_{\mathbb{R}^3}\mathcal{G}_r(\|\bm{y}\|)\bm{u}(\bm{x}-\bm{y},t)\,\textrm{d}\bm{y}$ denotes $\bm{u}$ filtered on the scale $r$, $\mathcal{G}_r$ is an isotropic kernel with filter length $r$, and 
${\bm{u}'\equiv \bm{u}-\widetilde{\bm{u}}}$ is the sub-grid field. Then, $\Delta\bm{u}= \Delta\widetilde{\bm{u}}+\Delta\bm{u}'$, and since $\widetilde{\bm{u}}$ is smooth on scales up to $\mathcal{O}(r)$ we may write $\Delta\widetilde{\bm{u}}(\bm{x},\bm{r},t)\approx \widetilde{\bm{\Gamma}}(\bm{x},t)\bm{\cdot r}$  \cite{li05}, where $\widetilde{\bm{\Gamma}}\equiv \bm{\nabla}\widetilde{\bm{u}}$ is the filtered velocity gradient. However, while $\bm{\partial_r\cdot}\Delta{\bm{u}}=0$ due to incompressibility, $\bm{\partial_r\cdot}\Delta\widetilde{\bm{u}}  \neq 0$,
since the filtering length is $r$. To avoid this compressibility issue
we instead define the filtered velocity increment in the solenoidal vector space
$\Delta^*\widetilde{\bm{u}}\equiv \bm{\partial_r}\times{\widetilde{\bm{\mathcal{A}}}^*}$, and using this we derive (see Appendix)
\begin{align}
\Delta^*\widetilde{\bm{u}} =\bm{\partial_r}\times \left[2\widetilde{\bm{\mathcal{A}}}(\bm{x}+\bm{r}/2,t)+2\widetilde{\bm{\mathcal{A}}}(\bm{x}-\bm{r}/2,t) +\widetilde{\bm{\mathcal{B}}}(\bm{x},t)\right],\label{II}
\end{align}
which satisfies $\bm{\partial_r\cdot}\Delta^*\widetilde{\bm{u}}(\bm{x},\bm{r},t)= 0$, where $\bm{\widetilde{\mathcal{A}}}$ is the vector potential defined through $\bm{\nabla}\times\widetilde{\bm{\mathcal{A}}}\equiv\bm{\widetilde{u}}$, and $\widetilde{\bm{\mathcal{B}}}$ is an integration constant. Next, we Taylor expand the $\widetilde{\bm{\mathcal{A}}}$ terms in Eq.~\eqref{II} in the variable $\bm{r}$. This is justified for two reasons: First, $\widetilde{\bm{\mathcal{A}}}$ is defined in terms of $\widetilde{\bm{u}}$, and $\widetilde{\bm{u}}$ is smooth at scales $\leq\mathcal{O}(r)$. Second, $\widetilde{\bm{\mathcal{A}}}$ is even smoother than  $\widetilde{\bm{u}}$ since $\widetilde{\bm{\mathcal{A}}}$ is given by the inverse curl operator (involving spatial integrals) acting on $\widetilde{\bm{u}}$. Terms up to second-order are explicitly retained, while the higher order terms are grouped into a remainder. Using this result we construct $\langle\Delta^*\widetilde{\bm{u}}\Delta^*\widetilde{\bm{u}}\Delta^*\widetilde{\bm{u}}\rangle$, and choose $\widetilde{\bm{\mathcal{B}}}$ so as to satisfy the incompressibility constraint \cite{hill97}
\begin{align}
\frac{\partial^3}{\partial  r_i\partial  r_j\partial  r_k}\langle\Delta^*\widetilde{u}_i\Delta^*\widetilde{u}_j \Delta^*\widetilde{u}_k\rangle=0.
\end{align}
Finally, by expressing $\langle\Delta^*\widetilde{\bm{u}}\Delta^*\widetilde{\bm{u}}\Delta^*\widetilde{\bm{u}}\rangle$ using its isotropic formula we obtain
\begin{eqnarray}
\begin{split}
\bm{\partial_r\cdot}\bm{T}&=\mathscr{L}\Big\{\langle (\widetilde{\bm{S}}\bm{\cdot}\widetilde{\bm{S}})\bm{:}\widetilde{\bm{S}}\rangle-\frac{1}{4}\langle\widetilde{\bm{\omega}}\widetilde{\bm{\omega}}\bm{:}\widetilde{\bm{S}}\rangle \Big\}+\mathcal{F},
\label{KRm1} 
\end{split}
\end{eqnarray}
where we have re-expressed $\widetilde{\bm{\Gamma}}$ in terms of the filtered strain-rate $\widetilde{\bm{S}}\equiv(\widetilde{\bm{\Gamma}}+\widetilde{\bm{\Gamma}}^\top)/2$, and the filtered vorticity $\widetilde{\bm{\omega}}\equiv\bm{\nabla}\times\widetilde{\bm{u}}$, and the operator $\mathscr{L}\{\cdot\}$ is defined as\[\mathscr{L}\{\cdot\}\equiv  (\partial_r+2/r) [(r^4/105)(\partial_r+7/r)\{\cdot\}].\]In Eq.~\eqref{KRm1}, $\mathcal{F}$ denotes the contributions involving both the sub-grid field $\Delta^*\bm{u}' \equiv\Delta{\bm{u}}-\Delta^*\widetilde{\bm{u}}$, and the higher order (sub-leading) terms in the expansion of $\widetilde{\bm{\mathcal{A}}}$. A detailed derivation of Eq.~\eqref{KRm1} is given in the Appendix. 

The general result in Eq.~\eqref{KRm1} that applies at all scales becomes more transparent when considering its asymptotic behavior in the dissipation and inertial ranges. In the limit  $r/\eta\to0$, $\mathcal{F}\to0$, and Eq.~\eqref{KRm1} reduces to
\begin{eqnarray}
\bm{\partial_r\cdot}\bm{T}= \frac{r^2}{3}\langle ({\bm{S}}\bm{\cdot}{\bm{S}})\bm{:}{\bm{S}}\rangle -\frac{r^2}{12}\langle{\bm{\omega}}{\bm{\omega}}\bm{:}{\bm{S}}\rangle,\label{KRdr}
\end{eqnarray}
where ${\bm{S}}= \lim_{r\to 0}\widetilde{\bm{S}}$ and ${\bm{\omega}}= \lim_{r\to 0}\widetilde{\bm{\omega}}$ are the ``bare''/un-filtered strain-rate and vorticity. In the inertial range, $\bm{\partial_r\cdot}\bm{T}$ is independent of $r$, which implies $ \langle (\widetilde{\bm{S}}\bm{\cdot}\widetilde{\bm{S}})\bm{:}\widetilde{\bm{S}}\rangle\propto \langle\widetilde{\bm{\omega}}\widetilde{\bm{\omega}}\bm{:}\widetilde{\bm{S}}\rangle\propto r^{-2}$, and using this in Eq.~\eqref{KRm1} we obtain 
for ${\eta\ll r\ll L}$ 
\begin{eqnarray}
\bm{\partial_r\cdot}\bm{T}=  \frac{r^2}{7}\langle (\widetilde{\bm{S}}\bm{\cdot}\widetilde{\bm{S}})\bm{:}\widetilde{\bm{S}}\rangle-\frac{r^2}{28}\langle\widetilde{\bm{\omega}}\widetilde{\bm{\omega}}\bm{:}\widetilde{\bm{S}}\rangle+\mathcal{F}. \label{KR}
\end{eqnarray}
It is anticipated that $\mathcal{F}$ will play a sub-leading role in Eq.~\eqref{KR} since numerical results in \cite{borue98} show that $\widetilde{\bm{u}}$ dominates the turbulent energy cascade as compared with $\bm{u}'$. This will be assumed in the following discussion, and later confirmed with data.

The result in Eq.~\eqref{KR} has similarities with the result obtained in \cite{eyink06} under the strong ultra-violet locality assumption (UVLA) for the instantaneous one-point scale-to-scale energy flux $\Pi(\bm{x},t)$, that describes the cascade of kinetic energy from $\widetilde{\bm{u}}$ to $\bm{u}'$. In contrast to $\Pi(\bm{x},t)$, however, the instantaneous form of $\bm{\partial_r\cdot}\bm{T}$, namely $(1/2)\bm{\partial_r\cdot}(\|\Delta\bm{u}\|^2\Delta\bm{u})$, is not in general reducible to a form such as Eq.~\eqref{KR}, even under UVLA. We also note that our general result in Eq.~\eqref{KRm1} differs from the energy flux result in \cite{eyink06} since our result in general depends on gradients in $r$-space of the strain and vorticity invariants. This difference arises since in \cite{eyink06} the multiscale properties of the turbulence are analyzed using a one-point field, whereas ours employs a two-point field representation in terms of the velocity increments. 

The result in Eq.~\eqref{KRm1} is in a sense kinematic, since it is essentially a re-expression of $\bm{\partial_r\cdot}\bm{T}$ in terms of the invariants of $\widetilde{\bm{\Gamma}}$. The dynamical information in Eq.~\eqref{KRm1} lies in the behavior of the invariants themselves. The invariant $(\widetilde{\bm{S}}\bm{\cdot}\widetilde{\bm{S}})\bm{:}\widetilde{\bm{S}}$ is the strain self-amplification (SSA) term, and contributes to production of $\|\widetilde{\bm{S}}\|$ when $(\widetilde{\bm{S}}\bm{\cdot}\widetilde{\bm{S}})\bm{:}\widetilde{\bm{S}}<0$. It is ``self'' production since it represents the excitation of the straining field by nonlinear interaction with itself. Written in the eigenframe of $\widetilde{\bm{S}}$ we have $(\widetilde{\bm{S}}\bm{\cdot}\widetilde{\bm{S}})\bm{:}\widetilde{\bm{S}}=\sum_i\widetilde{\lambda}_i^3$, where $\widetilde{\lambda}_1,\widetilde{\lambda}_2,\widetilde{\lambda}_3$ are the eigenvalues of $\widetilde{\bm{S}}$, satisfying ${\sum_i\widetilde{\lambda}_i=0}$ and $\widetilde{\lambda}_1\geq\widetilde{\lambda}_2\geq\widetilde{\lambda}_3$. It is known that $\langle (\widetilde{\bm{S}}\bm{\cdot}\widetilde{\bm{S}})\bm{:}\widetilde{\bm{S}}\rangle<0$ \cite{meneveau11}, and since $\widetilde{\lambda}_1\geq 0$, and $\langle\widetilde{\lambda}_2^3\rangle>0$ (see Fig.~\ref{Eig_cont}),
then $\langle (\widetilde{\bm{S}}\bm{\cdot}\widetilde{\bm{S}})\bm{:}\widetilde{\bm{S}}\rangle<0$ is solely due to the negativity of $\widetilde{\lambda}_3$. Taken together with Eq.~\eqref{KR}, this provides the conceptually appealing view that the contribution of SSA to the energy cascade is associated with compressional straining motion in the flow.

The invariant ${\widetilde{\bm{\omega}}\widetilde{\bm{\omega}}\bm{:}\widetilde{\bm{S}}}$ is the vortex stretching (VS) term, and when ${\widetilde{\bm{\omega}}\widetilde{\bm{\omega}}\bm{:}\widetilde{\bm{S}}>0}$ it contributes to the production of enstrophy $\|\widetilde{\bm{\omega}}\|^2$ through the stretching of (filtered) vortex lines. Written in the eigenframe, $\widetilde{\bm{\omega}}\widetilde{\bm{\omega}}\bm{:}\widetilde{\bm{S}}=\|\widetilde{\bm{\omega}}\|^2\widetilde{\lambda}_i\cos^2(\widetilde{\bm{\omega}},\widetilde{\bm{e}}_i)\equiv \widetilde{\mathcal{W}}_i$, where $\widetilde{\bm{e}}_i$ is the eigenvector corresponding to $\widetilde{\lambda}_i$. It is known that $\langle\widetilde{\bm{\omega}}\widetilde{\bm{\omega}}\bm{:}\widetilde{\bm{S}}\rangle>0$ \cite{meneveau11}, whose positivity can only come from $\widetilde{\lambda}_1$ or $\widetilde{\lambda}_2$. A well-known feature of turbulence is the predominant alignment of $\widetilde{\bm{\omega}}$ with $\widetilde{\bm{e}}_2$ \cite{meneveau11,danish18}. Nevertheless, the contribution to VS associated with $\widetilde{\lambda}_1$ dominates \cite{tsinober,doan18}. 

Since $\langle (\widetilde{\bm{S}}\bm{\cdot}\widetilde{\bm{S}})\bm{:}\widetilde{\bm{S}}\rangle<0$ and $\langle\widetilde{\bm{\omega}}\widetilde{\bm{\omega}}\bm{:}\widetilde{\bm{S}}\rangle>0$, then according to Eqs.~\eqref{KRdr} and \eqref{KR}, both the SSA and VS contribute to the downscale energy cascade. Note that here we are simply taking $\langle (\widetilde{\bm{S}}\bm{\cdot}\widetilde{\bm{S}})\bm{:}\widetilde{\bm{S}}\rangle<0$ and $\langle\widetilde{\bm{\omega}}\widetilde{\bm{\omega}}\bm{:}\widetilde{\bm{S}}\rangle>0$ as empirical facts. A complete explanation of the physics of the turbulent energy cascade would of course require an explanation for why these average invariants have the sign that they do. Several arguments have previously been given, however, all of them appear to be at best incomplete (e.g. \cite{tsinober}), and we do not attempt to provide new arguments. Nevertheless, independent of the explanation for why $\langle (\widetilde{\bm{S}}\bm{\cdot}\widetilde{\bm{S}})\bm{:}\widetilde{\bm{S}}\rangle<0$ and $\langle\widetilde{\bm{\omega}}\widetilde{\bm{\omega}}\bm{:}\widetilde{\bm{S}}\rangle>0$, the interpretation of these empiricial facts is unambiguous, namely, that nonlinearity in the NSE leads to the spontaneous production of strain and vorticity across the scales of the turbulent flow. Our goal is to understand how this spontaneous nonlinear production of strain and vorticity relates to the turbulent energy cascade.

Three additional points concerning our result for the energy cascade in Eq.~\eqref{KRm1} deserve further comment. The first is that our result does not assume locality of interactions among the scales of motion. Indeed, the filtered fields $\widetilde{\bm{S}}$ and $\widetilde{\bm{\omega}}$ involve contributions from all scales in the flow that are greater than or equal to the filter scale $r$, while the effects of the sub-grid scales are fully contained within $\mathcal{F}$. As a result, terms such as $\widetilde{\bm{\omega}}\widetilde{\bm{\omega}}\bm{:}\widetilde{\bm{S}}$ can involve contributions from interactions between scales of different sizes. Therefore, our theoretical results are consistent with recent numerical results that show that the stretching of vortices at a given scale tends to be governed by straining motions in the flow at scales that are 3 to 5 times larger \cite{doan18}.

The second is that in general, the average energy cascade depends, to leading order, on both the SSA and VS invariants, and also upon the gradients of these invariants in $r$-space. In the dissipation range $\partial_r \langle (\widetilde{\bm{S}}\bm{\cdot}\widetilde{\bm{S}})\bm{:}\widetilde{\bm{S}}\rangle=\partial_r \langle \widetilde{\bm{\omega}}\widetilde{\bm{\omega}}\bm{:}\widetilde{\bm{S}}\rangle=0$, because the invariants become independent of $r$ in this range. Outside of this range, if SSA and VS are scale invariant in $r$, then their gradients can be expressed in terms of the invariants themselves, e.g. if $\langle (\widetilde{\bm{S}}\bm{\cdot}\widetilde{\bm{S}})\bm{:}\widetilde{\bm{S}}\rangle\propto r^{-\xi}$ then $\partial_r \langle (\widetilde{\bm{S}}\bm{\cdot}\widetilde{\bm{S}})\bm{:}\widetilde{\bm{S}}\rangle\propto -\xi r^{-1}\langle (\widetilde{\bm{S}}\bm{\cdot}\widetilde{\bm{S}})\bm{:}\widetilde{\bm{S}}\rangle$. This is why in the inertial range result Eq.~\eqref{KR}, we were able to express the energy cascade solely in terms of the SSA and VS invariants. However, for situations where SSA and VS are not scale invariant (e.g. for low Reynolds number turbulence, or at the crossover between the dissipation and inertial ranges), then the energy cascade will depend to leading order on the gradients of SSA and VS in $r$-space, and not only on SSA and VS themselves.

The third is that according to our result, the invariant of the velocity gradient that is associated with the energy cascade is
\begin{align}
\begin{split}
\mathcal{R}^{\#}\equiv- (\widetilde{\bm{S}}\bm{\cdot}\widetilde{\bm{S}})\bm{:}\widetilde{\bm{S}}+\frac{1}{4}\widetilde{\bm{\omega}}\widetilde{\bm{\omega}}\bm{:}\widetilde{\bm{S}},
\end{split}
\end{align}
which differs from the $\mathcal{R}$ invariant typically studied in the context of the velocity gradient dynamics, namely \cite{chong90,cantwell93,meneveau11}
\begin{align}
\begin{split}
\mathcal{R}\equiv-\frac{1}{3} (\widetilde{\bm{S}}\bm{\cdot}\widetilde{\bm{S}})\bm{:}\widetilde{\bm{S}}-\frac{1}{4}\widetilde{\bm{\omega}}\widetilde{\bm{\omega}}\bm{:}\widetilde{\bm{S}}.
\end{split}
\end{align}
One implication of this is that the $\mathcal{Q},\mathcal{R}$ invariant maps (where ${\mathcal{Q}\equiv-(1/2)\bm{\nabla \widetilde{u} : \nabla \widetilde{u}}}$) usually employed in analyzing the dynamics of the fluid velocity gradients \cite{meneveau11} do not provide direct information concerning the energy transfer dynamics in turbulence (though they are of course related). A similar point to this was made in \cite{borue98}.
\subsection{Relative contributions of SSA and VS to the energy cascade} 
Betchov \cite{betchov56} derived results for the invariants of the velocity gradient tensor in statistically homogeneous turbulence, and extending this to the filtered fields we obtain 
\begin{eqnarray}
\begin{split}
\langle (\widetilde{\bm{S}}\bm{\cdot}\widetilde{\bm{S}})\bm{:}\widetilde{\bm{S}}\rangle=-\frac{3}{4}\langle\widetilde{\bm{\omega}}\widetilde{\bm{\omega}}\bm{:}\widetilde{\bm{S}}\rangle,\,\forall r.\label{BRcg}
\end{split}
\end{eqnarray}
It is important to stress that this result is purely kinematic/statistical; it is derived without reference to NSE, and can be proven assuming only incompressibility and statistical homogeneity of the flow.

Using Eq.~\eqref{BRcg} we observe that the contribution from SSA in Eqs.~\eqref{KRdr} and \eqref{KR} is three times larger than that from VS, indicating that VS is not the main driver of the energy cascade, and that it in fact plays a sub-leading role. SSA is the main mechanism driving the energy cascade. 
\subsection{Correctly distinguishing the roles of SSA and VS in the energy cascade} 
One may object to the conclusion that SSA, not VS, dominates the energy cascade since if we substitute Eq.~\eqref{BRcg} into \eqref{KR} we obtain
\begin{eqnarray}
\begin{split}
\bm{\partial_r\cdot T}=-\frac{r^2}{7}\langle\widetilde{\bm{\omega}}\widetilde{\bm{\omega}}\bm{:}\widetilde{\bm{S}}\rangle+\mathcal{F},\label{KR2}
\end{split}
\end{eqnarray}
(a similar step to this was taken in \cite{eyink06b}, namely, the Betchov relation in Eq.~\eqref{BRcg} was used to express the energy flux purely in terms of VS for a homogeneous turbulent flow) which appears to show that VS is the mechanism governing the downscale energy cascade, contrary to our previous statements. Indeed, Eq.~\eqref{KR2} would explain why previous numerical studies (e.g. \cite{davidson08,doan18}) seemed to find a strong correlation between the energy cascade and VS, on average. 

Nevertheless, we argue that Eq.~\eqref{KR2} is fundamentally misleading with respect to the physical mechanism driving the energy cascade (and therefore so also is Eq. (32) in \cite{eyink06b}). Namely, it invokes Eq.~\eqref{BRcg} which is a purely kinematic relationship that obscures the fact that SSA and VS are dynamically very different. That SSA and VS are dynamically very different, and therefore must be correctly distinguished, may be observed in at least three different ways. First is the simple fact that the dynamical equations governing the SSA and VS invariants that can be derived from the NSE are very different (see appendix C of \cite{tsinober}). Second, SSA and VS can affect the evolution of other quantities in turbulent flows in completely different ways. For example, consider the equation for $\langle\|\widetilde{\bm{S}}\|^2\rangle$ (for a homogeneous flow)
\begin{align}
\begin{split}
\frac{1}{2}\partial_t \langle\|\widetilde{\bm{S}}\|^2\rangle&= -\langle(\widetilde{\bm{S}}\bm{\cdot}\widetilde{\bm{S}})\bm{:}\widetilde{\bm{S}}\rangle -\frac{1}{4}\langle\widetilde{\bm{\omega}}\widetilde{\bm{\omega}}\bm{:}\widetilde{\bm{S}}\rangle-\nu\langle\|\bm{\nabla \widetilde{S}}\|^2\rangle.\\
\end{split}
\label{S2eq}
\end{align}
From Eq.~\eqref{S2eq} it is apparent that while $\langle\widetilde{\bm{\omega}}\widetilde{\bm{\omega}}\bm{:}\widetilde{\bm{S}}\rangle>0$ acts as a \emph{sink} for $\langle\|\widetilde{\bm{S}}\|^2\rangle$, $\langle(\widetilde{\bm{S}}\bm{\cdot}\widetilde{\bm{S}})\bm{:}\widetilde{\bm{S}}\rangle<0$ acts as a \emph{source} for $\langle\|\widetilde{\bm{S}}\|^2\rangle$. Therefore, while $\langle (\widetilde{\bm{S}}\bm{\cdot}\widetilde{\bm{S}})\bm{:}\widetilde{\bm{S}}\rangle=-(3/4)\langle\widetilde{\bm{\omega}}\widetilde{\bm{\omega}}\bm{:}\widetilde{\bm{S}}\rangle$, the negativity of  $\langle (\widetilde{\bm{S}}\bm{\cdot}\widetilde{\bm{S}})\bm{:}\widetilde{\bm{S}}\rangle$ leads to an opposite dynamical effect on $ \langle\|\widetilde{\bm{S}}\|^2\rangle$ than the negativity of $-\langle\widetilde{\bm{\omega}}\widetilde{\bm{\omega}}\bm{:}\widetilde{\bm{S}}\rangle$. Third, while the average values of SSA and VS are closely related, their statistics in general differ, and joint Probability Density Functions (PDFs) of the bare SSA and VS reveal a weak correlation coefficient of $\approx 0.26$ \cite{gulitski07} (the results in \S\ref{NRD} reveal something similar for the filtered fields).


These arguments emphasize that while inserting $\langle (\widetilde{\bm{S}}\bm{\cdot}\widetilde{\bm{S}})\bm{:}\widetilde{\bm{S}}\rangle=-(3/4)\langle\widetilde{\bm{\omega}}\widetilde{\bm{\omega}}\bm{:}\widetilde{\bm{S}}\rangle$ into Eq.~\eqref{KR} is numerically legitimate, it obscures the true physics behind the energy cascade because VS and SSA are distinct dynamical processes that have distinct effects on the dynamics of turbulence. Their roles in the cascade mechanism must therefore be distinguished; the true underlying physics is reflected in Eq.~\eqref{KR}, not \eqref{KR2}.

The above arguments are analogous to the argument that even though in homogeneous turbulence, $\langle\epsilon\rangle\equiv 2\nu\langle \|\bm{S}\|^2\rangle=\nu\langle\|\bm{\omega}\|^2\rangle$, it is dynamically incorrect to refer to $\nu\langle\|\bm{\omega}\|^2\rangle$ as the average dissipation rate, since vorticity has no direct causal relationship with dissipation \cite{tennekes}. Indeed, for an incompressible, Newtonian fluid, $\epsilon\equiv2\nu\|\bm{S}\|^2$, \emph{by definition}. The result $\langle\epsilon\rangle=\nu\langle\|\bm{\omega}\|^2\rangle$, like $\langle (\widetilde{\bm{S}}\bm{\cdot}\widetilde{\bm{S}})\bm{:}\widetilde{\bm{S}}\rangle=-(3/4)\langle\widetilde{\bm{\omega}}\widetilde{\bm{\omega}}\bm{:}\widetilde{\bm{S}}\rangle$, is purely kinematic, and must not be interpreted as implying any actual dynamical or causal relationship between the quantities involved.

For completeness, we also mention that in \cite{davidson} the following result is derived for $r\to0$
\begin{eqnarray}
\begin{split}
\langle\Delta u_\parallel^3(r,t)\rangle=-\frac{2}{35}\langle{\bm{\omega}}{\bm{\omega}}\bm{:}{\bm{S}}\rangle r^3,\label{Dav}
\end{split}
\end{eqnarray}
where $\Delta u_\parallel\equiv r^{-1}\bm{r\cdot} \Delta\bm{u}$ is the component of $\Delta\bm{u}$ parallel to $\bm{r}$. However, we would argue that one should not interpret this result as establishing a dynamical connection between VS and the skewness of $\Delta u_\parallel$. Indeed, for $r\to0, \Delta u_\parallel=r^{-1}\bm{rr :S}$, so that vorticity plays no explicit role and so should not explicitly appear in the expression for $\langle\Delta u_\parallel^3(r,t)\rangle$. For an isotropic flow we can derive the result 
\begin{eqnarray}
\begin{split}
\langle\Delta u_\parallel^3(r,t)\rangle= \langle r^{-3}(\bm{rr :S})^3 \rangle  =\frac{8}{105}\langle ({\bm{S}}\bm{\cdot}{\bm{S}})\bm{:}{\bm{S}} \rangle r^3,\quad\text{for}\,\,r\to0,\label{DavC}
\end{split}
\end{eqnarray}
which shows that the real relationship is between SSA and the skewness of $\Delta u_\parallel$, rather than between VS and the skewness of $\Delta u_\parallel$. Application of Eq.~\eqref{BRcg} transforms Eq.~\eqref{DavC} into Eq.~\eqref{Dav}, but in doing so confuses which dynamical invariant is associated with $\langle\Delta u_\parallel^3(r,t)\rangle$. The reason why Eq.~\eqref{Dav} is derived in \cite{davidson} rather than Eq.~\eqref{DavC} is precisely because their derivation employs the Betchov relation \eqref{BRcg} to re-write expressions involving SSA in terms of VS, and in doing so, conflates their dynamical roles.

\section{Mechanisms governing energy dissipation and their relation to the mechanisms of the energy cascade}\label{EDEC}
%
%
In \cite{taylor37}, Taylor \& Green argue that in isotropic, non-stationary turbulence, the mechanism of VS leads to an increase in the average kinetic energy dissipation rate. On the other hand, in \cite{tsinober,sagaut18} it is argued that VS leads to a suppression of the dissipation rate, and that this in turn implies that VS must hinder the energy cascade, since dissipation is supposed to be the end result of the cascade. In this section we resolve these contradictory assertions.

The argument of \cite{taylor37} is that if VS leads to an increase of vorticity in the flow, then this leads to an increase in the quantity $\nu\langle\|\bm{\omega}\|^2\rangle$ which they refer to as the average kinetic energy dissipation rate. However, as mentioned earlier, this is incorrect. By definition, for a Newtonian fluid the average dissipation is $\langle\epsilon\rangle\equiv 2\nu\langle\|\bm{S}\|^2\rangle$, not $\nu\langle\|\bm{\omega}\|^2\rangle$. Now if VS leads to an increase of $\langle\|\bm{\omega}\|^2\rangle$, then since for a homogeneous flow we have the kinematic result $\langle\epsilon\rangle=\nu\langle\|\bm{\omega}\|^2\rangle$, then VS will \emph{indirectly} lead to an increase of $\langle\epsilon\rangle$. However, this connection is purely kinematic (indeed, for an inhomogeneous flow, $\langle\|\bm{\omega}\|^2\rangle\neq  2\langle\|\bm{S}\|^2\rangle$ so that an increase in the value of $\langle\|\bm{\omega}\|^2\rangle$ caused by VS need not correspond numerically to an increase in the value of $\langle\epsilon\rangle$). In a strict, direct, dynamical sense, VS leads to a decrease, not increase, in the value of $\langle\epsilon\rangle$. This may be seen by considering the equation governing $\langle\|{\bm{S}}\|^2\rangle$ (and therefore $\langle\epsilon\rangle$) for a homogeneous flow 

\begin{align}
\begin{split}
\frac{1}{2}\partial_t\langle\|{\bm{S}}\|^2\rangle=& -\langle({\bm{S}}\bm{\cdot}{\bm{S}})\bm{:}{\bm{S}}\rangle -\frac{1}{4}\langle{\bm{\omega}}{\bm{\omega}}\bm{:}{\bm{S}}\rangle-\nu\langle\|\bm{\nabla {S}}\|^2\rangle.\\
\end{split}
\label{S2eqb}
\end{align}
This result clearly shows that when $\langle{\bm{\omega}}{\bm{\omega}}\bm{:}{\bm{S}}\rangle>0$, $\langle\|{\bm{S}}\|^2\rangle$, and therefore $\langle\epsilon\rangle$, is suppressed. Therefore, in a strict dynamical sense, VS actually acts to reduce the dissipation rate, not increase it. The nonlinear dynamical process responsible for enhancing the dissipation is SSA. This is essentially the argument put forward in \cite{tsinober,sagaut18}. 

While we therefore agree with \cite{tsinober} and \cite{sagaut18} in arguing that VS dynamically opposes dissipation, we think their assertion that this implies that VS hinders the energy cascade is incorrect. Indeed, our analytical results show that VS contributes to the downscale cascade, and does not hinder it. But this appears paradoxical; how can VS contribute to the downscale cascade of energy, while at the same time acting to reduce the dissipation?

We suggest that the argument in \cite{tsinober,sagaut18} involves a confusion concerning the nature of the connection between the energy cascade and energy dissipation. The interpretation of the inertial range result $\bm{\partial_r\cdot}\bm{T}=-2\langle\epsilon\rangle$ is that in the stationary state, $\partial_t\mathcal{K}=0$, there is a balance between the energy received by scale $r$ due to $\bm{\partial_r\cdot}\bm{T}$, and the rate at which energy is passed down to smaller scales, which is equal to $-2\langle\epsilon\rangle$. Therefore, the result $\bm{\partial_r\cdot}\bm{T}=-2\langle\epsilon\rangle$ does not mean that the mechanism of $\bm{\partial_r\cdot}\bm{T}$ is the dynamical cause of the energy dissipation, but rather it simply reflects an energetic balance between the two processes. Namely, the energy passed down by the cascade is given to the small scales where it is then dissipated. This point can be made clearer by considering that for statistically stationary, homogeneous turbulence, $\mathcal{P}=\langle\epsilon\rangle$, where $\mathcal{P}$ is the kinetic energy production term \cite{pope}. According to this result, there is a balance between the energy injected into the flow by the production mechanism, and the energy dissipated. However, this does not imply that the production is the dynamical cause of dissipation. Indeed, the way in which kinetic energy is produced in the flow is somewhat arbitrary (e.g. it may be by a mean-shear in the flow, or injection of energy by random stirring etc), while the energy dissipation is governed by NSE. In view of these considerations, there is no reason why VS has to contribute dynamically to $\bm{\partial_r\cdot}\bm{T}$ and $-2\langle\epsilon\rangle$ in the same way, since $\bm{\partial_r\cdot}\bm{T}=-2\langle\epsilon\rangle$ implies an energetic balance not a dynamical, causal relationship. Therefore, there is no actual paradox in the assertion that VS contributes to the downscale energy cascade, while at the same time acting to reduce the dissipation rate.

\section{Numerical results \& Discussion}\label{NRD}
We now turn to test the theoretical results using data from a Direct Numerical Simulation (DNS) of the incompressible NSE. The NSE are solved using the Highly Parallel Particle-laden flow Solver for Turbulence Research (HiPPSTR) psuedo-spectral code \cite{ireland13} on a triply-periodic domain of length $2\pi$ with $2048^3$ grid points. The Taylor Reynolds number of the statistically steady and isotropic flow is $R_\lambda=597$. Further details on the DNS code and simulations may be found in \cite{ireland13,ireland16a}. A sharp-spectral filter was used to construct $\widetilde{\bm{\omega}}$ and $\widetilde{\bm{S}}$ for use in Eq.~\eqref{KRm1}, but we also compared the results to those obtained using a Gaussian filter (see \cite{pope} for details on filter methods), and found similar results.

%
\FloatBarrier

\begin{figure}[h]
	\centering
	\vspace{0mm}
	{\begin{overpic}
			[trim = 0mm 0mm 0mm 0mm,scale=1.0,clip,tics=20]{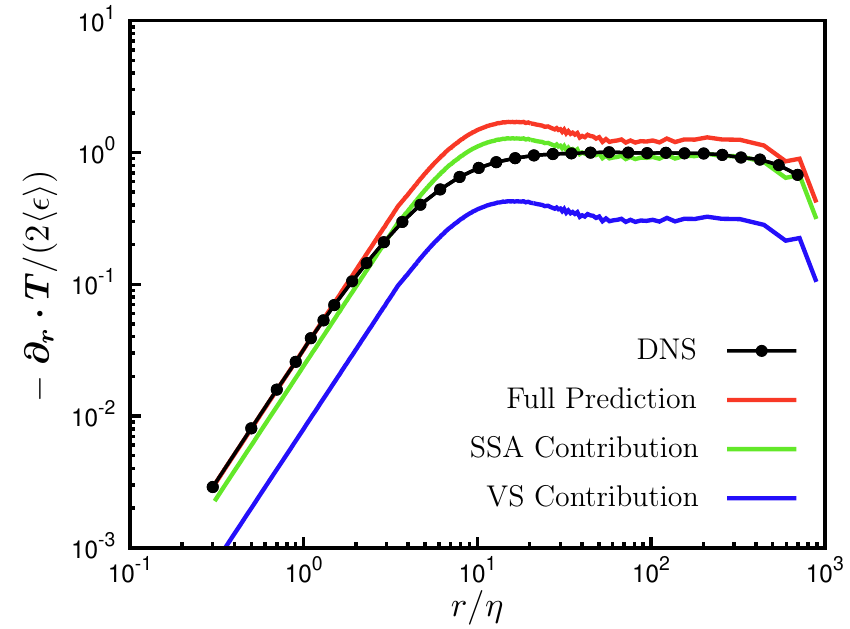}
	\end{overpic}}
	\caption{Comparison of DNS data (black line with circles) for $\bm{\partial_r\cdot T}$ with Eq.~\eqref{KRm1} (red line). Also shown are the SSA (green line) and VS (blue line) contributions to Eq.~\eqref{KRm1}}
	\label{divT}
\end{figure}
\FloatBarrier
In Fig.~\ref{divT} we compare the DNS results for $\bm{\partial_r\cdot T}$ with the rhs of Eq.~\eqref{KRm1}. Since we do not know $\mathcal{F}$ we set it to zero when plotting the results. In the limit $r/\eta \to0$ this introduces no approximation since $\lim_{r/\eta\to0}\mathcal{F}\to0$. The DNS data confirms the accuracy of this asymptotic behavior even up to $r=\mathcal{O}(\eta)$. The results in Fig.~\ref{divT} imply that in the inertial range, $\mathcal{F}$ makes a finite, but sub-leading contribution to the cascade.

We also plot separately the SSA and VS contributions to Eq.~\eqref{KRm1}. The results confirm that in both the dissipation and inertial ranges, the energy cascade is dominated by the contribution from SSA rather than VS, with the contribution from SSA being three times larger than that from VS. These results therefore confirm the prediction of our analysis that across the range of scales in a turbulent flow, the stretching of vorticity plays a sub-leading role. The self-amplification of the strain-rate field dominates the energy cascade. 

\begin{figure}[h]
	\centering
	\vspace{0mm}
	\subfloat[]
	{\begin{overpic}
			[trim = 0mm 0mm 0mm 0mm,scale=1.0,clip,tics=20]{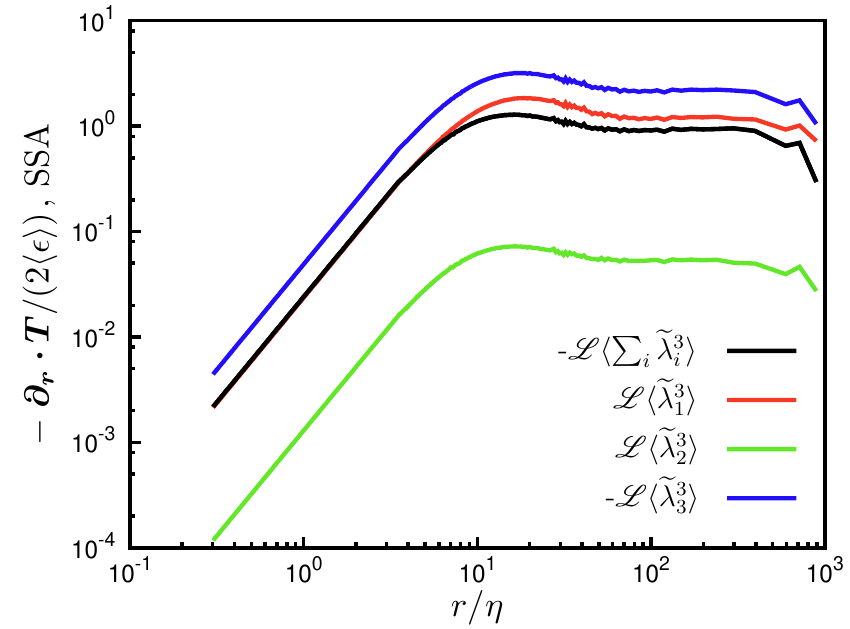}
	\end{overpic}}
		\subfloat[]
		{\begin{overpic}
			[trim = 0mm 0mm 0mm 0mm,scale=1.0,clip,tics=20]{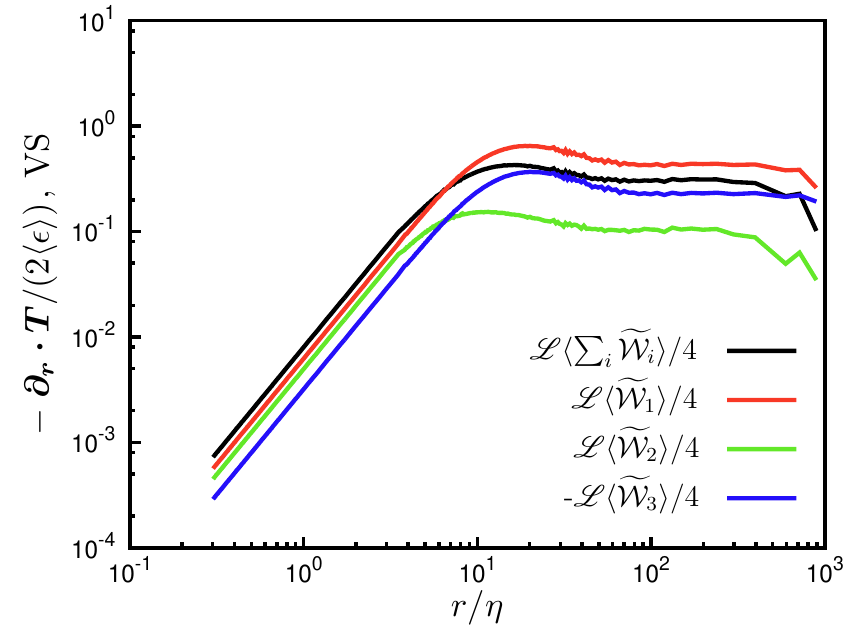}
	\end{overpic}}
	\caption{Eigenframe contributions from (a) SSA and (b) VS to  Eq.~\eqref{KRm1}.}
	\label{Eig_cont}
\end{figure}

In Fig.~\ref{Eig_cont}(a) we show the DNS data for the eigenframe contributions to $\mathscr{L}\langle\sum_i \widetilde{\lambda}_i^3\rangle$, where ${(\widetilde{\bm{S}}\bm{\cdot}\widetilde{\bm{S}})\bm{:}\widetilde{\bm{S}}=\sum_i\widetilde{\lambda}_i^3}$, and the results show that the $i=3$ contribution dominates and is the sole cause of the negativity of $\mathscr{L}\langle\sum_i \widetilde{\lambda}_i^3\rangle$ at all scales in the flow. This shows that the SSA process that dominates the energy cascade is itself governed by compressional straining motions at all scales.
\FloatBarrier

\noindent In Fig.~\ref{Eig_cont}(b) we plot the eigenframe contributions to $\mathscr{L}\langle\sum_i \widetilde{\mathcal{W}}_i\rangle$, where ${\widetilde{\bm{\omega}}\widetilde{\bm{\omega}}\bm{:}\widetilde{\bm{S}}=\sum_i\|\widetilde{\bm{\omega}}\|^2\widetilde{\lambda}_i\cos^2(\widetilde{\bm{\omega}},\widetilde{\bm{e}}_i)\equiv\sum_i \widetilde{\mathcal{W}}_i}$, and the results show that the contribution from $i=1$ is the most positive at all scales. This is consistent with the results in \cite{doan18} that show that at all scales in the flow, VS is dominated by the contribution from the extensional eigenvalue. Interestingly, while our results show $\langle\widetilde{\mathcal{W}}_1\rangle>\langle\widetilde{\mathcal{W}}_2\rangle$ in the dissipation range (as observed in \cite{gulitski07}), $\langle\widetilde{\mathcal{W}}_1\rangle\gg\langle\widetilde{\mathcal{W}}_2\rangle$ in the inertial range. While $\widetilde{\mathcal{W}}_1$ and $\widetilde{\mathcal{W}}_3$ have fixed signs, the sign of $\widetilde{\mathcal{W}}_2$ fluctuates, and as a result, the contribution to ${\sum_i\langle\widetilde{\mathcal{W}}_i\rangle}$ from $\langle\widetilde{\mathcal{W}}_2\rangle$ may be smaller than that from $\langle\widetilde{\mathcal{W}}_1\rangle$ due to partial cancellation of positive and negative $\widetilde{\mathcal{W}}_2$ in its average. To explore this, we computed $\langle|\widetilde{\mathcal{W}}_2|\rangle$ and found that
at all scales,
$\langle\widetilde{\mathcal{W}}_2\rangle<\langle|\widetilde{\mathcal{W}}_2|\rangle<\langle\widetilde{\mathcal{W}}_1\rangle$ (although our data indicates $\lim_{r\to 0}\langle|\widetilde{\mathcal{W}}_2|\rangle/\langle\widetilde{\mathcal{W}}_1\rangle\to 1$), such that the dominance of the $i=1$ contribution to ${\sum_i\langle\widetilde{\mathcal{W}}_i\rangle}$ is not simply caused by the fluctuating sign of $\widetilde{\mathcal{W}}_2$. It is mainly because $\widetilde{\lambda}_1$ tends to be larger than $|\widetilde{\lambda}_2|$, so that $\langle\widetilde{\mathcal{W}}_1\rangle$
dominates ${\sum_i\langle\widetilde{\mathcal{W}}_i\rangle}$, despite the fact that $\widetilde{\bm{\omega}}$ preferentially aligns with $\widetilde{\bm{e}}_2$ at all scales in the flow \cite{danish18}.

\begin{figure}[h]
	\centering
	\vspace{0mm}
	\subfloat[]
	{\begin{overpic}
			[trim = 0mm 0mm 0mm 0mm,scale=1.0,clip,tics=20]{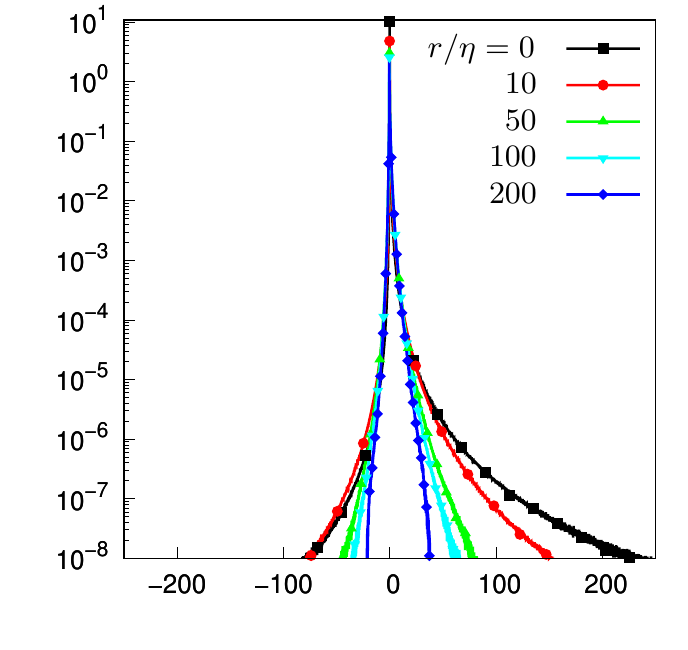}
			\put(95,2){\large$(\xi-\mu)/\sigma$}	
			\put(-0,90){\large\rotatebox{90}{$\sigma$PDF}}						
	\end{overpic}}
	\subfloat[]
	{\begin{overpic}
			[trim = 0mm 0mm 0mm 0mm,scale=1.0,clip,tics=20]{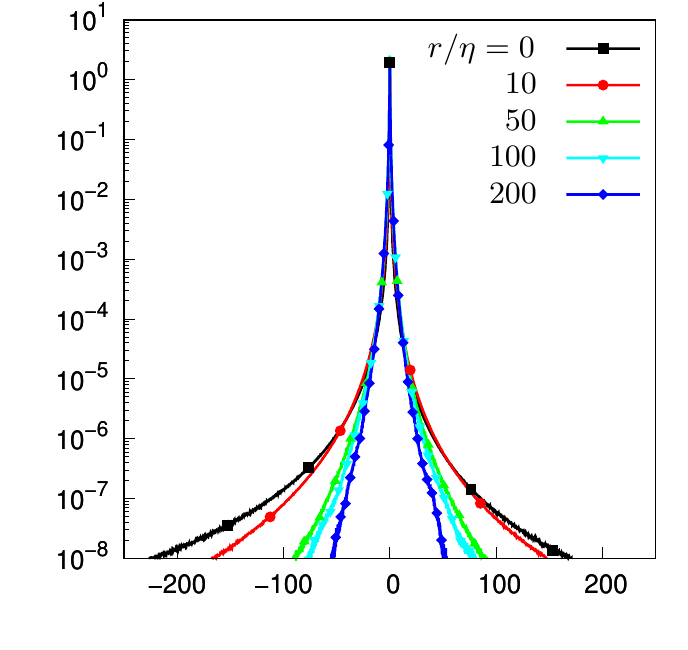}
			\put(95,2){\large$(\xi-\mu)/\sigma$}	
			\put(-0,90){\large\rotatebox{90}{$\sigma$PDF}}								
	\end{overpic}}	
	\caption{DNS results for the PDFs of (a) $\xi=\mathcal{R}^{\#}$ and (b) $\xi=\mathcal{R}$, normalized by their respective mean $\mu$ and standard deviation $\sigma$, at different filter scales $r/\eta$.}
	\label{PDF_RsR}
\end{figure}
\FloatBarrier
Having considered the average energy cascade, which depends on the average values of the SSA and VS invariants, we now turn to consider the fluctuations in the SSA and VS terms which can give insight into the behavior of the energy cascade during fluctuations away from its mean behavior.

We begin by considering the PDF of $\mathcal{R}^{\#}\equiv- (\widetilde{\bm{S}}\bm{\cdot}\widetilde{\bm{S}})\bm{:}\widetilde{\bm{S}}+(1/4)\widetilde{\bm{\omega}}\widetilde{\bm{\omega}}\bm{:}\widetilde{\bm{S}}$ (whose mean value determines the average energy cascade in \eqref{KRm1}) measured at different filtering scales $r/\eta$. The results in figure~\ref{PDF_RsR}(a) show that not only is the mean value of the variable positive (associated with the energy cascade being downscale on average), but it is also strongly positively skewed. This implies that the probability of events associated with a downscale transfer of energy is much larger than for upscale transfer events. As the filter scale $r/\eta$ is increased, the mean and skewness decreases, reflecting the fact that the production of strain and vorticity becomes weaker, and less intermittent,  as one moves to larger scales where the velocity gradients are themselves weaker. For comparison, we also plot the PDFs of the invariant $\mathcal{R}\equiv-(1/3) (\widetilde{\bm{S}}\bm{\cdot}\widetilde{\bm{S}})\bm{:}\widetilde{\bm{S}}-(1/4)\widetilde{\bm{\omega}}\widetilde{\bm{\omega}}\bm{:}\widetilde{\bm{S}}$ that is commonly considered in studies of turbulence dynamics \cite{chong90,cantwell93,meneveau11}. For homogeneous turbulence, the mean value of $\mathcal{R}$ is zero according to the Betchov relation \eqref{BRcg}. We also see that compared with $\mathcal{R}^{\#}$, the PDF of $\mathcal{R}$ is more symmetric. This difference is mainly due to the difference in sign of the contribution of VS to $\mathcal{R}^{\#}$ and $\mathcal{R}$. This difference is important, however, since it is $\mathcal{R}^{\#}$, not $\mathcal{R}$, that is the dynamical invariant associated with the energy cascade.

\begin{figure}[h]
	\centering
	\vspace{0mm}
	{\begin{overpic}
			[trim = 0mm 0mm 0mm 0mm,scale=1.0,clip,tics=20]{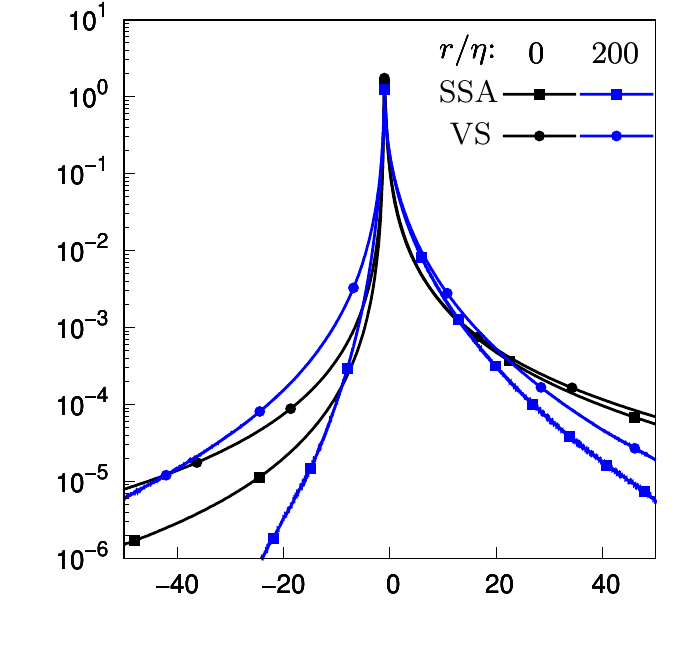}
			\put(105,2){\large$\xi/\mu $}	
			\put(-0,90){\large\rotatebox{90}{$\mu$PDF}}						
	\end{overpic}}
	\caption{DNS results for the PDFs of the variable $\xi$ (either SSA - lines with squares, or VS - lines with circles), normalized by the mean value $\mu$, and at different filter scales $r/\eta$.}
	\label{PDF_SSA_VS}
\end{figure}
\FloatBarrier

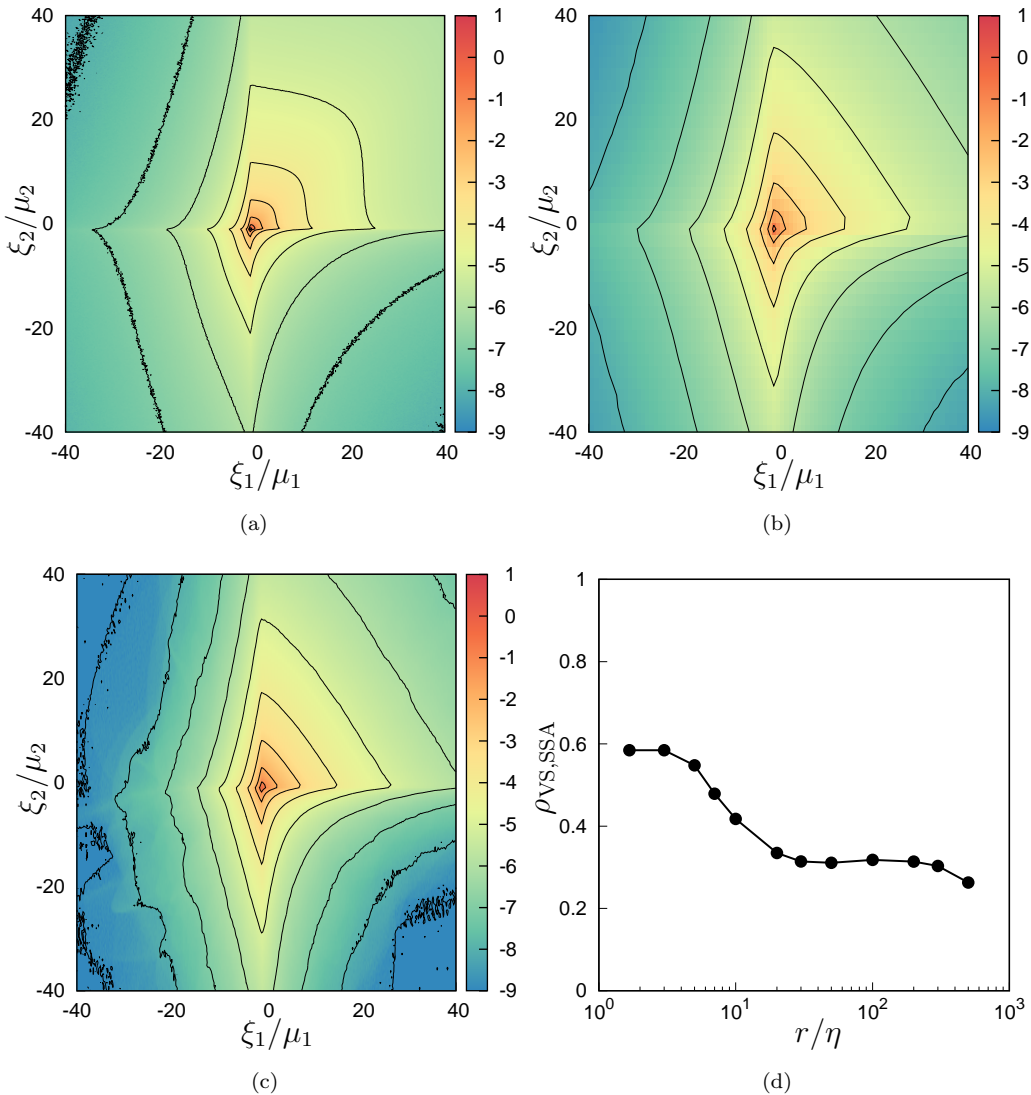
\begin{figure}[h]
	\centering
	\vspace{0mm}
	\subfloat[]
	{\begin{overpic}
			[trim = 0mm 0mm 0mm 0mm,scale=1.0,clip,tics=20]{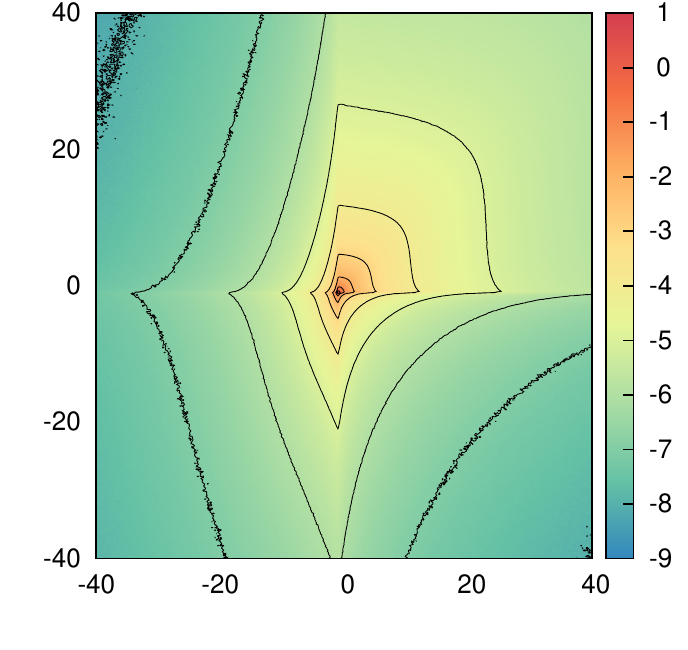}
			\put(90,7){\large$\xi_1/\mu_1$}	
			\put(5,93){\large\rotatebox{90}{$\xi_2/\mu_2$}}							
	\end{overpic}}
		\subfloat[]
	{\begin{overpic}
			[trim = 0mm 0mm 0mm 0mm,scale=1.0,clip,tics=20]{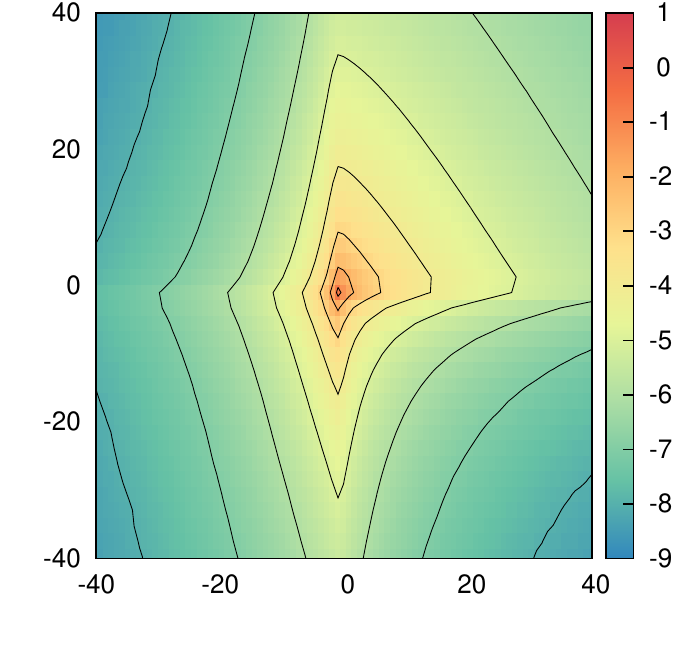}
			\put(90,7){\large$\xi_1/\mu_1$}	
			\put(5,93){\large\rotatebox{90}{$\xi_2/\mu_2$}}									
	\end{overpic}}\\
		\subfloat[]
	{\begin{overpic}
			[trim = 0mm 0mm 0mm 0mm,scale=1.0,clip,tics=20]{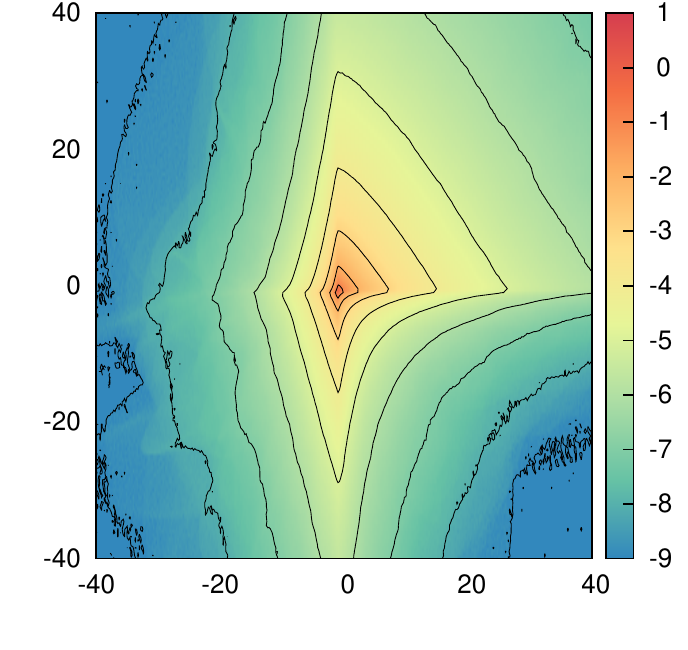}
			\put(90,7){\large$\xi_1/\mu_1$}	
			\put(5,93){\large\rotatebox{90}{$\xi_2/\mu_2$}}									
	\end{overpic}}
			\subfloat[]
	{\hspace{-3mm}\begin{overpic}
			[trim = 0mm 0mm 0mm 0mm,scale=1.0,clip,tics=20]{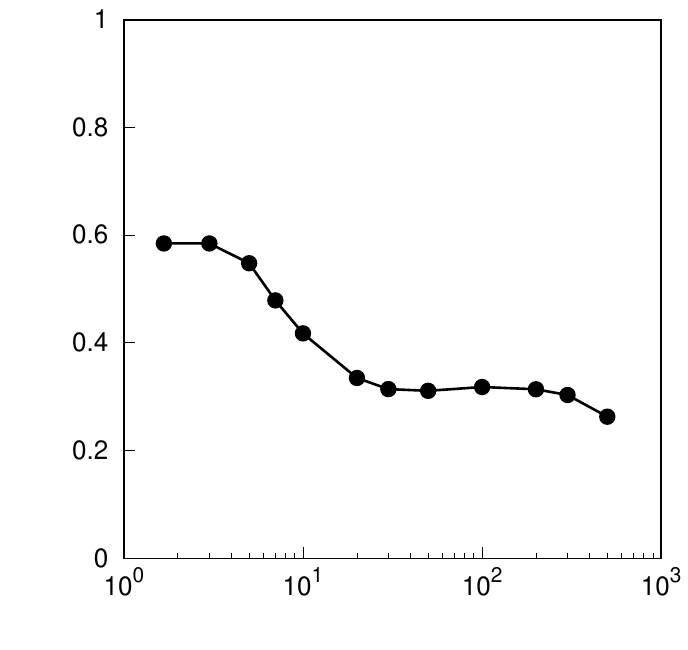}
			\put(110,7){\large$r/\eta$}	
			\put(12,93){\large\rotatebox{90}{$\rho_{\textrm{VS,SSA}}$}}								
	\end{overpic}}
	\caption{DNS results for the joint PDF of SSA and VS at filter scales (a) $r/\eta=0$, (b) $r/\eta=50$, (c) $r/\eta=200$. The quantity on the horizontal axis corresponds to SSA divided by its mean, and the vertical axis corresponds to VS divided by its mean. The colors correspond to $\log_{10}$PDF. Plot (d) is the correlation coefficient for VS and SSA as a function of filter scale.}
	\label{jPDF_SSA_VS}
\end{figure}

In Figure~\ref{PDF_SSA_VS} we show results for the normalized PDFs of SSA and VS at different filtering scales. In contrast with the results in \cite{gulitski07} where it was shown that the PDFs of the unfiltered SSA and VS are almost identical, we find significant differences, between the PDFs (there are several possible reasons for the discrepancy between our results and those of \cite{gulitski07}, including that their experimental results are for the atmospheric surface layer, not isotropic turbulence). Most noticeably, the probability of events corresponding to vortex compression (i.e. $\widetilde{\bm{\omega}}\widetilde{\bm{\omega}}\bm{:}\widetilde{\bm{S}}<0$) is significantly larger than that for events corresponding to the self destruction of the strain field (i.e. $(\widetilde{\bm{S}}\bm{\cdot}\widetilde{\bm{S}})\bm{:}\widetilde{\bm{S}}>0$). For $r/\eta=200$, corresponding to the inertial range, the difference in the SSA and VS PDFs becomes even more apparent. These results provide further support for our earlier assertions that the roles of SSA and VS must not be conflated. Not only are they dynamically very different processes, but furthermore, as Figure~\ref{PDF_SSA_VS} shows, the general statistics of the two processes are significantly different, despite the fact that their mean values are closely related through \eqref{BRcg}. This point is made even clearer in Figure~\ref{jPDF_SSA_VS}(a)-(c) where we show results for the joint PDF of VS and SSA at different filtering scales. The shape of the PDF contours in Figure~\ref{jPDF_SSA_VS}(a)-(c) are similar to those in the experimental work \cite{gulitski07} (they only considered the unfiltered case), showing the distinctive ``corners" of the contour lines along $\xi_1/\mu_1=0$ and $\xi_2/\mu_2=0$. Figure 5(d) shows the correlation coefficient of VS and SSA, $\rho_{\textrm{VS,SSA}}$. While $\rho_{\textrm{VS,SSA}}$ is moderate in the dissipation range, in the inertial range $\rho_{\textrm{VS,SSA}}$ approaches $0.3$, indicating a weak correlation.


\FloatBarrier

\begin{figure}[h]
	\centering
	\vspace{0mm}
	\subfloat[]
	{\begin{overpic}
			[trim = 5mm 0mm 0mm 0mm,scale=1.0,clip,tics=20]{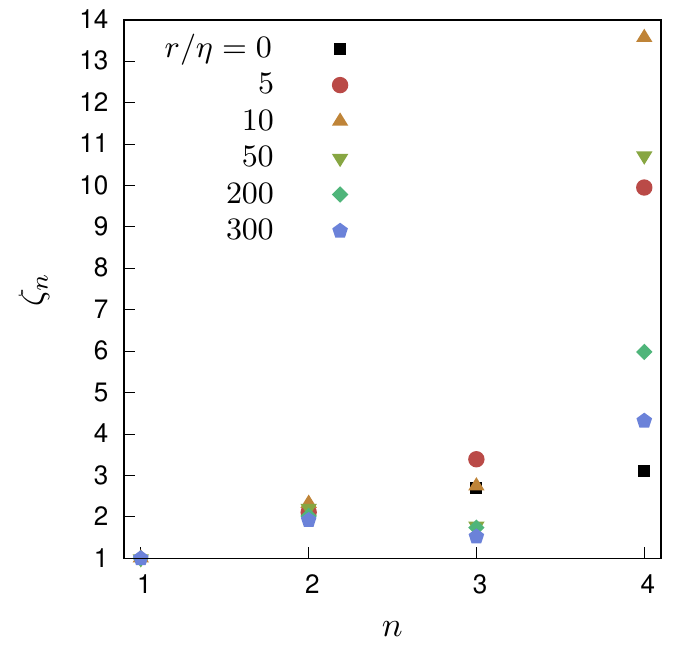}
			\put(-2,100){\large\rotatebox{90}{$\zeta_n$}}							
	\end{overpic}}
	\subfloat[]
	{\begin{overpic}
			[trim = 5mm 0mm 0mm 0mm,scale=1.0,clip,tics=20]{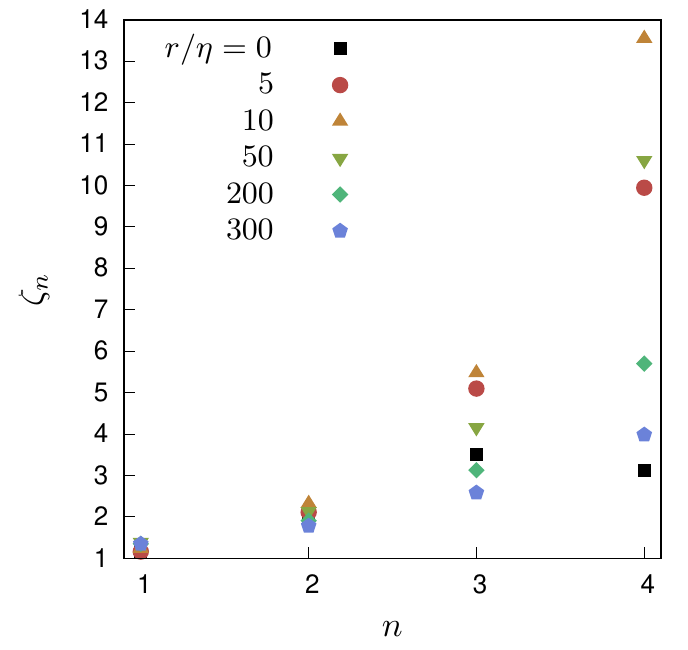}
			\put(-2,100){\large\rotatebox{90}{$\zeta_n^{\textrm{abs}}$}}	
	\end{overpic}}
	\caption{DNS results for (a) $\zeta_n\equiv \langle[(-3/4)\widetilde{\bm{\omega}}\widetilde{\bm{\omega}}\bm{:}\widetilde{\bm{S}}]^n\rangle/\langle[ (\widetilde{\bm{S}}\bm{\cdot}\widetilde{\bm{S}})\bm{:}\widetilde{\bm{S}}]^n\rangle$ and (b) $\zeta_n^{\textrm{abs}}\equiv \langle|(-3/4)\widetilde{\bm{\omega}}\widetilde{\bm{\omega}}\bm{:}\widetilde{\bm{S}}|^n\rangle/\langle| (\widetilde{\bm{S}}\bm{\cdot}\widetilde{\bm{S}})\bm{:}\widetilde{\bm{S}}|^n\rangle$ at different filtering scales $r/\eta$.}
	\label{zetaN}
\end{figure}

Finally, we have argued that concerning the average energy cascade, the contribution from SSA is much larger than that from VS. However, it is important to consider whether the same holds true for fluctuations of the energy cascade about its average behavior. To explore this we consider the quantity
\begin{eqnarray}
\begin{split}
\zeta_n\equiv \frac{\langle[(-3/4)\widetilde{\bm{\omega}}\widetilde{\bm{\omega}}\bm{:}\widetilde{\bm{S}}]^n\rangle}{\langle[ (\widetilde{\bm{S}}\bm{\cdot}\widetilde{\bm{S}})\bm{:}\widetilde{\bm{S}}]^n\rangle}.\label{GBRcg}
\end{split}
\end{eqnarray}
The results for $\zeta_n$ are shown in Figure~\ref{zetaN}(a) for different $n$ and filtering scales $r/\eta$. For $n=1$, \eqref{BRcg} gives $\zeta_1=1$, as observed in our numerical results for each $r/\eta$. However, for $n>1$, $\zeta_n>1$ at each scale, and reaches values $\mathcal{O}(10)$ for $n=4$, implying that VS may play a leading order role in the energy cascade during strong fluctuations of the cascade about its average value. This increasingly important role of VS compared with SSA during large fluctuations of the energy cascade about the average behavior may in part be associated with the known fact that the vorticity field is more intermittent that the strain-rate field in turbulent flows \cite{donzis08,chen97,buaria19}. 

\FloatBarrier

Two further points concerning the results in Figure~\ref{zetaN}(a) deserve comment. First, for a given $n$, $\zeta_n$ depends non-monotonically on $r/\eta$, and is maximum at $r/\eta=\mathcal{O}(10)$ for each $n$ considered. Second, at some scales, $\zeta_n$ depends non-monotonically on $n$. To check the cause of this, we also computed
\begin{eqnarray}
\begin{split}
\zeta_n^{\textrm{abs}}\equiv \frac{\langle|(-3/4)\widetilde{\bm{\omega}}\widetilde{\bm{\omega}}\bm{:}\widetilde{\bm{S}}|^n\rangle}{\langle| (\widetilde{\bm{S}}\bm{\cdot}\widetilde{\bm{S}})\bm{:}\widetilde{\bm{S}}|^n\rangle},\label{GBRcgb}
\end{split}
\end{eqnarray}
which involves the absolute values of the invariants. The results in Figure~\ref{zetaN}(b) show that $\zeta_n^{\textrm{abs}}$ does not suffer from this non non-monotonicity. Therefore, the likely explanation of the behavior of $\zeta_n$ is that unlike even moments, odd moments involve cancellation of terms in the average, so that even and odd moments may behave differently.

\section{Conclusions} 
In this paper we have considered, using theoretical and numerical approaches, the mechanisms governing the turbulent energy cascade. Our results show that the traditionally invoked mechanism of vortex stretching (VS) is not the main contributor to the average energy cascade in isotropic turbulence. Instead, the main mechanism driving the energy cascade is the strain self-amplification (SSA) process, and this remains true in both the dissipation and inertial ranges of the turbulence. Central to our results and arguments is our assertion regarding the need to carefully distinguish between the dynamical roles of VS and SSA in the cascade. Despite some (limited) similarities in the statistics of these processes in homogeneous turbulence, VS and SSA correspond to very different dynamical processes, and so their dynamical effects on the cascade must not be conflated. We have also provided an argument to show that the fact that VS contributes to the downscale energy cascade, while inhibiting the fluid kinetic energy dissipation rate are not contradictory assertions. This apparent contradiction is resolved by noting that the balance between the flux of energy through the inertial range and the energy dissipation rate only reflects a statistical, energetic balance, not a dynamical or causal relationship between the two processes.

Using DNS data we also showed that at every scale in the flow, the contribution of SSA to the average downscale energy cascade is exclusively due to the behavior of the compressional eigenvalue of the filtered strain-rate tensor. On the other hand, the contribution of VS to the average downscale energy cascade comes from both the extensional and intermediate eigenvalues of the filtered strain-rate tensor. However, while the contribution to VS from the extensional eigenvalue is only slightly larger than that from the intermediate eigenvalue in the dissipation range, in the inertial range, the contribution to VS from the extensional eigenvalue is much larger than that from the intermediate eigenvalue.

Finally, our DNS results also indicate that while VS plays a sub-leading role in the average energy cascade, it may play a leading order role during large fluctuations of the energy cascade about its average behavior. This seems to be connected to the well known fact that the vorticity field is more intermittent than the strain-rate field in turbulence.

As a next step, we plan to extend the analysis using the $\mathrm{SO}(3)$ group to analyze flows of lower symmetry, such as axisymmetric turbulent flows involving additional forces and effects on the flow.
\section{Acknowledgements}
The authors would like to acknowledge G.\ Katul, A.\ Porporato, and M. Iovieno for stimulating discussions on the topic. This work used the Extreme Science and Engineering Discovery Environment (XSEDE), supported by National Science Foundation grant ACI-1548562 \cite{xsede}.

\section{Appendix}

In this appendix we provide details of the steps in the analysis leading to the analytical results presented in the paper. 

We wish to evaluate the nonlinear energy flux $\bm{\partial_r\cdot T}$ at scale $r$. The aim is to relate $\bm{\partial_r\cdot T}$ to vortex stretching at the same scale $r$, which can be achieved by considering the stretching of the filtered vorticity field by the filtered strain-rate field. To this end we first introduce $\bm{u}=\widetilde{\bm{u}}+\bm{u}'$, where 
\begin{align}
\widetilde{\bm{u}}(\bm{x},t)\equiv\int_{\mathbb{R}^3}\mathcal{G}_r(\|\bm{y}\|)\bm{u}(\bm{x}-\bm{y},t)\,\mathrm{d}\bm{y},
\end{align}
denotes $\bm{u}$ filtered on the scale $r$, $\mathcal{G}_r$ is an isotropic kernel with filter length $r$, and ${\bm{u}'\equiv \bm{u}-\widetilde{\bm{u}}}$ is the sub-grid field. Then we decompose the fluid velocity increment $\Delta\bm{u}$ as $\Delta\bm{u}= \Delta\widetilde{\bm{u}}+\Delta\bm{u}'$, where $\Delta\widetilde{\bm{u}}\equiv\widetilde{\bm{u}}(\bm{x}+\bm{r}/2,t)-\widetilde{\bm{u}}(\bm{x}-\bm{r}/2,t)$, and $\Delta\bm{u}'\equiv\Delta\bm{u}= \Delta\widetilde{\bm{u}}$. However, while $\bm{\partial_r\cdot}\Delta{\bm{u}}=0$ due to incompressibility, $\bm{\partial_r\cdot}\Delta\widetilde{\bm{u}}\neq 0$ and $\bm{\partial_r\cdot}\Delta{\bm{u}}'\neq 0$, since the filtering length defining $\widetilde{\bm{u}}$ is $r$. In order to avoid this compressibility issue we must use a more sophisticated approach that allows us to construct a filtered velocity increment that satisfies the incompressibility condition. We now show how this is achieved.

\subsection{Incompressible filtered fluid velocity increment}

Let us first define the more general filtered field
\begin{align}
\widetilde{\bm{u}}(\bm{x},t)\equiv\int_{\mathbb{R}^3}\mathcal{G}_\ell(\|\bm{y}\|)\bm{u}(\bm{x}-\bm{y},t)\,\mathrm{d}\bm{y},
\end{align}
where now the filtering lengthscale is $\ell$ which is at this stage arbitrary. Then we, will denote the incompressible filtered fluid velocity increment by $\Delta^*\widetilde{\bm{u}}(\bm{x},\bm{r},\ell,t)$, where $\ell$ denotes the scale on which the velocity field has been filtered. A constraint on the definition of $\Delta^*\widetilde{\bm{u}}$ is that it should reduce to the traditional increment $\Delta\widetilde{\bm{u}}$ when the filtering length $\ell$ is held fixed (i.e. independent of $r$), since then $\bm{\partial_r\cdot}\Delta\widetilde{\bm{u}}= 0$ because $\bm{\partial_r\cdot}$ commutes with the filtering operator when $\ell$ is independent of $r$. In view of these we define
\begin{align}
\frac{\partial}{\partial r_i} \Delta^*\widetilde{u}_i(\bm{x},\bm{r},\ell,t)\Big\vert_{\bm{x}} &= 0,\label{div_du}\\
\Delta^*\widetilde{u}_i(\bm{x},\bm{r},\ell,t)\Big\vert_{\ell} &= \Delta\widetilde{u}_i(\bm{x},\bm{r},\ell,t),\label{congr}
\end{align}
where $\vert_{\bm{x}}$, $\vert_{\ell}$ denote that the quantity is evaluated at fixed $\bm{x}$ and $\ell$, respectively.

The formal solution of Eq.~\eqref{div_du} can be expressed as
\begin{align}
\Delta^*\widetilde{u}_i(\bm{x},\bm{r},\ell,t)= \epsilon_{ijk} \frac{\partial}{\partial r_j} \widetilde{\mathcal{A}}^*_k(\bm{x},\bm{r},\ell,t)\Big\vert_{\bm{x}},\label{sol_du}
\end{align}
%
where $\widetilde{\bm{\mathcal{A}}}^*$ is a vector potential of the incompressible increment. Substituting this expression for $\Delta^*\widetilde{\bm{u}}$ into Eq.~\eqref{congr} we obtain
\begin{align}
\epsilon_{ijk}\frac{\partial}{\partial r_j} \widetilde{\mathcal{A}}^*_k(\bm{x},\bm{r},\ell,t)\Big\vert_{\bm{x},\ell} = \Delta\widetilde{u}_i(\bm{x},\bm{r},\ell,t).\label{bar_du}
\end{align}
This now clarifies the notation used in Eq.~\eqref{congr}; $\Delta^*\bm{\widetilde{u}}(\bm{x},\bm{r},\ell,t)\vert_{\ell}$ means that when, according to the vector potential expression for $\Delta^*\bm{\widetilde{u}}$, the curl operator $\bm{\partial_r}\times\{\cdot\}$ acts on $\widetilde{\bm{\mathcal{A}}}^*$, $\ell$ is to be held fixed.

The solution of Eq.~\eqref{bar_du} reads
\begin{align}
\widetilde{\mathcal{A}}^*_k(\bm{x},\bm{r},\ell,t) = 2{\widetilde{\mathcal{A}}}_k(\bm{x}+\bm{r}/2,\ell,t) + 2{\widetilde{\mathcal{A}}}_k(\bm{x}-\bm{r}/2,\ell,t) + \widetilde{\mathcal{B}}_k(\bm{x},\ell,t)\label{sol_G}
\end{align}
where ${\widetilde{\bm{\mathcal{A}}}}$ is the vector potential associated with the velocity field filtered at length $\ell$, defined through,
\begin{align}
\widetilde{u}_i(\bm{x},\ell,t) = \epsilon_{ijk}\frac{\partial}{\partial x_j}{\widetilde{\mathcal{A}}}_k(\bm{x},\ell,t)\Big\vert_{\ell},\label{def1}
\end{align}
and that we used in the following way:
\begin{align}
\widetilde{u}_i(\bm{x}+\bm{r}/2,\ell,t) = 2\epsilon_{ijk}\frac{\partial }{\partial r_j}{\widetilde{\mathcal{A}}}_k(\bm{x}+\bm{r}/2,\ell,t)\Big\vert_{\bm{x},\ell}.\label{def1}
\end{align}
The term $\bm{\widetilde{\mathcal{B}}}(\bm{x},\ell,t)$ in Eq.~\eqref{sol_G} is an integration constant, since we are integrating Eq.~\eqref{bar_du} with respect to $\bm{r}$ at constant $\bm{x}$ and $\ell$. We then obtain the expression for the incompressible filtered velocity increment by substituting Eq.~\eqref{sol_G} into \eqref{sol_du} and evaluating the increment at a scale-dependent filtering length:
\begin{align}
\Delta^*\widetilde{u}_i(\bm{x},\bm{r},\ell,t) = \epsilon_{ijk} \frac{\partial}{\partial r_j}\Big(2{\widetilde{\mathcal{A}}}_k(\bm{x}+\bm{r}/2,\ell,t) + 2{\widetilde{\mathcal{A}}}_k(\bm{x}-\bm{r}/2,\ell,t) + \widetilde{\mathcal{B}}_k(\bm{x},\ell,t)\Big)\Big\vert_{\bm{x}}.\label{IFVIa}
\end{align}
Note that using Eq.~\eqref{def1} we can re-write this in the more illuminating form
\begin{align}
\begin{split}
\Delta^*\widetilde{u}_i(\bm{x},\bm{r},\ell,t) = &\Delta\widetilde{u}_i(\bm{x},\bm{r},\ell,t)  \\
&+2\epsilon_{ijk}\frac{r_j}{r}\frac{\textrm{d}\ell}{\textrm{d}r} \frac{\partial}{\partial \ell}\Big({\widetilde{\mathcal{A}}}_k(\bm{x}+\bm{r}/2,\ell,t) + {\widetilde{\mathcal{A}}}_k(\bm{x}-\bm{r}/2,\ell,t) + \frac{1}{2}\widetilde{\mathcal{B}}_k(\bm{x},\ell,t)\Big)\Big\vert_{\bm{x}}.
\end{split}
\label{IFVI}
\end{align}
In order to relate $\Delta^*\bm{\widetilde{u}}(\bm{x},\bm{r},\ell,t)$ to the velocity gradient filtered at scale $r$, namely $\widetilde{\bm{\Gamma}}\equiv \bm{\nabla}\widetilde{\bm{u}}$, we choose $\ell=r$ and Taylor expand the terms involving $\widetilde{\bm{\mathcal{A}}}$ in Eq.~\eqref{IFVIa} in the variable $\bm{r}$. This is justified for two reasons: First, $\widetilde{\bm{\mathcal{A}}}$ is defined in terms of $\widetilde{\bm{u}}$, and $\widetilde{\bm{u}}$ is smooth at scales $\leq\mathcal{O}(r)$ since we use $\ell=r$. Second, $\widetilde{\bm{\mathcal{A}}}$ is even smoother than  $\widetilde{\bm{u}}$ since $\widetilde{\bm{\mathcal{A}}}$ is given by the inverse curl operator (involving spatial integrals) acting on $\widetilde{\bm{u}}$. 

The expansion may be written as
\begin{align}
{\widetilde{\mathcal{A}}}_k(\bm{x}+\bm{r}/2,\ell,t) +{\widetilde{\mathcal{A}}}_k(\bm{x}-\bm{r}/2,\ell,t) =2{\widetilde{\mathcal{A}}}_k(\bm{x},\ell,t)+\frac{1}{4}r_p r_q\frac{\partial^2}{\partial x_p\partial x_q} {\widetilde{\mathcal{A}}}_k(\bm{x},\ell,t)+h_k,\label{Aexp}\\
h_k(\bm{x},\bm{r},\ell,t) \equiv {\widetilde{\mathcal{A}}}_k(\bm{x}+\bm{r}/2,\ell,t) +{\widetilde{\mathcal{A}}}_k(\bm{x}-\bm{r}/2,\ell,t) -2{\widetilde{\mathcal{A}}}_k(\bm{x},\ell,t)-\frac{1}{4}r_p r_q\frac{\partial^2}{\partial x_p\partial x_q} {\widetilde{\mathcal{A}}}_k(\bm{x},\ell,t),\label{h}
\end{align}
where $\bm{h}$ denotes the higher-order (fourth order and higher) terms in the expansion of $\bm{\widetilde{\mathcal{A}}}$. Due to the smoothness of $\bm{\widetilde{\mathcal{A}}}$ on scales $\leq\mathcal{O}(r)$, $\bm{h}$ will be a sub-leading contribution to $\bm{\widetilde{\mathcal{A}}}$ in Eq.~\eqref{Aexp}. However, since the expansion of $\bm{\widetilde{\mathcal{A}}}$ is likely only asymptotic, we define $\bm{h}$ via Eq.~\eqref{h} rather than by the explicit summation of the higher order terms in the Taylor series expansion, which could be divergent.

Substituting Eqs.~\eqref{Aexp} and \eqref{h} into \eqref{IFVIa}, and using $\ell=r$ together with the definition of $\bm{\widetilde{\mathcal{A}}}$, we obtain 
\begin{align}
\begin{split}
\Delta^* \widetilde{u}_i(\bm{x},\bm{r},\ell=r,t) = &\widetilde{\Gamma}_{ij}(\bm{x},r,t)r_j + \epsilon_{ijk}\frac{r_j}{r} \left(\frac{1}{2}\frac{\partial^3\widetilde{\mathcal{A}}_k}{\partial r \partial x_p\partial x_q}(\bm{x},r,t)r_pr_q + \frac{\partial \widetilde{\mathfrak{B}}_k}{\partial r}(\bm{x},r,t)\right)\\
&+ 2\epsilon_{ijk}\frac{\partial }{\partial r_j}h_k(\bm{x},\bm{r},\ell=r,t).
\end{split}
\label{IFVIT}
\end{align}
To write the second term in the form shown we have used the fact that $\bm{\widetilde{\mathcal{A}}}(\bm{x},r,t)$ and $\bm{\widetilde{\mathfrak{B}}}(\bm{x},r,t)\equiv\bm{\widetilde{\mathcal{B}}}(\bm{x},r,t)+ 4\bm{\widetilde{\mathcal{A}}}(\bm{x},r,t)$ are functions of $r$, but do not depend upon the orientation of the vector $\bm{r}$. In contrast, $\bm{h}$ depends on the full vector $\bm{r}$ and so cannot be written this way.

The second terms in Eqs.~\eqref{IFVI} and \eqref{IFVIT} represent the ``compressible correction'' that captures the effect of the variable filtering length on the velocity increment, and guarantee that $\bm{\partial_r\cdot}\Delta^*\bm{\widetilde{u}}(\bm{x},\bm{r},\ell,t)=0\forall\ell$. For example, considering Eq.~\eqref{IFVIT}, while $\bm{\partial_r\cdot}(\bm{\partial_r}\times\bm{h})=0$, the divergence of the first and second terms in Eq.~\eqref{IFVIT} are both non zero, but precisely cancel each other out, ensuring that $\bm{\partial_r\cdot}\Delta^*\bm{\widetilde{u}}=0$.

Note that in the paper, when referring to $\widetilde{\bm{\mathcal{A}}}^*$, $\widetilde{\bm{\mathcal{A}}}$ and $\widetilde{\bm{\mathcal{B}}}$, for notational clarity, we do not explicitly write the $r$ dependence in the argument of these tensor functions. This dependence is signified implicitly through the $\widetilde{\cdot}$ notation.

\subsection{Constructing the energy cascade expression}
The second term on the rhs of Eqs.~\eqref{IFVI} and \eqref{IFVIT} lies in the plane orthogonal to the vector $\bm{r}$. As such, the longitudinal increments satisfy $\Delta^* \widetilde{u}_\parallel=\Delta \widetilde{u}_\parallel$, while the perpendicular increments differ, $\Delta^* \widetilde{u}_\perp\neq\Delta \widetilde{u}_\perp$. This fact allows us to greatly simply the construction of the result for $\bm{\partial_r\cdot T}$
using Eq.~\eqref{IFVIT}.

Using the results in \cite{hill97}, since $\bm{\partial_r\cdot}\Delta^*\bm{\widetilde{u}}(\bm{x},\bm{r},\ell,t)\vert_{\bm{x}}=0$ we have the constraint
\begin{align}
\frac{\partial^3}{\partial  r_i\partial  r_j\partial  r_k}\langle\Delta^*\widetilde{u}_i\Delta^*\widetilde{u}_j \Delta^*\widetilde{u}_k\rangle=0.
\end{align}
Then, as shown in \cite{hill97}, for an isotropic flow we can write the constraint above as
\begin{align}
\langle\Delta^*\widetilde{u}_\parallel\Delta^*\widetilde{u}_\perp^2 \rangle-\frac{1}{6}\Bigg(\langle\Delta^*\widetilde{u}_\parallel^3\rangle+r\frac{\partial}{\partial r} \langle\Delta^*\widetilde{u}_\parallel^3\rangle\Bigg)=0.\label{Constraint}
\end{align}
Now, from the definition of $\Delta^*\bm{\widetilde{u}}(\bm{x},\bm{r},\ell,t)$ in Eq.~\eqref{IFVI} we have
\begin{align}
\langle\Delta^*\widetilde{u}_\parallel^3\rangle&=\langle\Delta\widetilde{u}_\parallel^3\rangle,\label{upar}\\
\langle\Delta^*\widetilde{u}_\parallel\Delta^*\widetilde{u}_\perp^2 \rangle&=\langle\Delta\widetilde{u}_\parallel\Delta\widetilde{u}_\perp^2 \rangle+C,\label{uparper}
\end{align}
where $C$ denotes the contribution arising from the compressibility correction term in \eqref{IFVI}. Instead of directly computing $C$ from Eq.~\eqref{IFVI}, which is very involved, 
we can instead determine it using Eqs.~\eqref{upar}, \eqref{uparper} together with \eqref{Constraint} yielding
\begin{align}
C=\frac{1}{6}\Bigg(\langle\Delta\widetilde{u}_\parallel^3\rangle+r\frac{\partial}{\partial r}  \langle\Delta\widetilde{u}_\parallel^3\rangle\Bigg)-\langle\Delta\widetilde{u}_\parallel\Delta\widetilde{u}_\perp^2 \rangle.\label{C}
\end{align}
Next, we introduce the decomposition $\Delta\bm{u}=\Delta^*\bm{\widetilde{u}}+\Delta^*\bm{{u}}'$ into the definition of $\bm{\partial_r\cdot T}$ to obtain
\begin{align}
\bm{\partial_r\cdot T}=(1/2)\bm{\partial_r\cdot}\langle\|\Delta^*\widetilde{\bm{u}}\|^2\Delta^*\widetilde{\bm{u}}\rangle+{F},
\end{align}
where ${F}$ denotes the contributions involving the sub-grid field $\Delta^*\bm{{u}}'$. For an isotropic flow we have
\begin{align}
\bm{\partial_r\cdot T}=\frac{1}{2}\Big({\partial_r}+\frac{2}{r} \Big)  \Big(\langle\Delta^*\widetilde{u}_\parallel^3\rangle+2 \langle\Delta^*\widetilde{u}_\parallel\Delta^*\widetilde{u}_\perp^2 \rangle\Big)  +{F}.\label{divTan}
\end{align}
%
Then, introducing Eq.~\eqref{IFVIT} into \eqref{upar} and \eqref{uparper}, and using the appropriate isotropic tensor formulas we obtain
\begin{align}
\langle\Delta^*\widetilde{u}_\parallel^3\rangle &=\langle  \widetilde{{\Gamma}}_{im}\widetilde{{\Gamma}}_{jn}\widetilde{{\Gamma}}_{kp}\rangle\frac{r_m r_n r_p r_i r_j r_k}{r^3}+\mathcal{H}_\parallel= \frac{2}{35} \langle (\widetilde{\bm{\Gamma}}\bm{\cdot}\widetilde{\bm{\Gamma}})\bm{:}\widetilde{\bm{\Gamma}}^\top\rangle r^3+\mathcal{H}_\parallel,\label{upar2}\\
\langle\Delta^*\widetilde{u}_\parallel\Delta^*\widetilde{u}_\perp^2 \rangle&=\langle  \widetilde{{\Gamma}}_{im}\widetilde{{\Gamma}}_{jn}\widetilde{{\Gamma}}_{kp}\rangle\frac{r_m r_n r_p r_i n_j n_k}{r n^2}+C+\mathcal{H}_\perp=\frac{4}{105} \langle (\widetilde{\bm{\Gamma}}\bm{\cdot}\widetilde{\bm{\Gamma}})\bm{:}\widetilde{\bm{\Gamma}}^\top\rangle r^3+C+\mathcal{H}_\perp,\label{uper2}
\end{align}
where $\mathcal{H}_\parallel$ and $\mathcal{H}_\perp$ denote the contributions arising from $\bm{h}$ in Eq.~\eqref{IFVIT}. Substituting these into Eq.~\eqref{divTan} and using \eqref{C} we finally obtain
\begin{align}
\begin{split}
\bm{\partial_r\cdot}\bm{T}&=\mathscr{L}\Big\{\langle (\widetilde{\bm{S}}\bm{\cdot}\widetilde{\bm{S}})\bm{:}\widetilde{\bm{S}}\rangle-\frac{1}{4}\langle\widetilde{\bm{\omega}}\widetilde{\bm{\omega}}\bm{:}\widetilde{\bm{S}}\rangle \Big\}+\mathcal{F},
\label{KRm1B} 
\end{split}
\end{align}
where we have re-expressed $\langle (\widetilde{\bm{\Gamma}}\bm{\cdot}\widetilde{\bm{\Gamma}})\bm{:}\widetilde{\bm{\Gamma}}^\top\rangle$ in terms of the filtered strain-rate $\widetilde{\bm{S}}\equiv(\widetilde{\bm{\Gamma}}+\widetilde{\bm{\Gamma}}^\top)/2$, and the filtered vorticity $\widetilde{\bm{\omega}}\equiv\bm{\nabla}\times\widetilde{\bm{u}}$, using the exact decomposition
\begin{align}
\begin{split}
\langle (\widetilde{\bm{\Gamma}}\bm{\cdot}\widetilde{\bm{\Gamma}})\bm{:}\widetilde{\bm{\Gamma}}^\top\rangle=\langle (\widetilde{\bm{S}}\bm{\cdot}\widetilde{\bm{S}})\bm{:}\widetilde{\bm{S}}\rangle-\frac{1}{4}\langle\widetilde{\bm{\omega}}\widetilde{\bm{\omega}}\bm{:}\widetilde{\bm{S}}\rangle,
\end{split}
\end{align}
the operator $\mathscr{L}\{\cdot\}$ is defined as\[\mathscr{L}\{\cdot\}\equiv  (\partial_r+2/r) [(r^4/105)(\partial_r+7/r)\{\cdot\}],\]and $\mathcal{F}$ is now the sum of $F$ and the contributions from $\mathcal{H}_\parallel$ and $\mathcal{H}_\perp$.

\bibliography{E_cascade_paper}

\end{document}